\documentclass[dvips]{article}
\usepackage{url}

\usepackage{icrc2011}
\usepackage{graphicx}
\usepackage{epstopdf}
\usepackage{url}
\title{Automated analysis of {\em Fermi}-LAT data to trigger ground-based gamma-ray observations}
\newcommand{\etal}{\MakeLowercase{\textit{et al. }}} % "et al."
\newcommand{\g}{$\gamma$ }
\newcommand{\F}{{\em Fermi} }
\newcommand{\FL}{{\em Fermi}-LAT }
\shorttitle{M. Errando \etal Automated {\em Fermi}-LAT analysis for IACT observations}

\authors{M. Errando$^{1}$, M. Orr$^{2}$, E. Kara$^{1}$}
\afiliations{$^1$Department of Physics and Astronomy, Barnard College, Columbia University, NY 10027, USA\\ 
$^2$ Department of Physics and Astronomy, Iowa State University, Ames, IA 50011, USA
}
\email{errando@astro.columbia.edu}

\abstract{The {\em Fermi} Gamma-ray Space Telescope regularly surveys the entire sky in the energy range between 0.3 and 100 GeV with an homogeneous coverage. This makes {\em Fermi} a very useful guide for ground-based Cherenkov-telescope arrays like VERITAS, that are sensitive at energies above $\sim$100 GeV. The VERITAS collaboration uses information from {\em Fermi}-LAT to select potential targets and identify flaring objects in the GeV band. We describe three different analysis pipelines that automatically process and analyze {\em Fermi}-LAT data on a daily basis to look for flaring objects: daily light curves of selected targets, an all-sky analysis to identify new flaring sources, and an algorithm that looks for clusters of high-energy photons in time and arrival direction.}

\begin{document}
\maketitle

%Begin the section.
\section{Introduction}

The {\em Fermi} Gamma-ray Space Telescope operates in survey mode giving a view of the entire sky in the energy range between 100\,MeV and 300\,GeV every 3\,h. The sensitivity of the Large Area Telescope (LAT)  onboard the {\em Fermi} satellite is good enough to detect bright $\gamma$-ray sources in time scales as short as one day. This is potentially interesting for ground-based $\gamma$-ray Cherenkov telescopes like VERITAS, that can use the all-sky monitoring capabilities of \F to identify flaring $\gamma$-ray sources in the GeV range that could potentially be detected at TeV energies from ground.

The \FL collaboration notifies the community about flaring $\gamma$-ray sources through Astronomer's Telegrams, blogs \cite{fermiblogday,fermiblogweek}, and publicly available light curves for a number of objects of interest \cite{fermilc}. However, the information is typically made public 1-3 days after the main flaring event happened, and sometimes the high gamma-ray activity is over. To reduce the latency time between the gamma-ray flare and potential VERITAS observations, and also to extend the Fermi monitoring to other objects that are of interest for TeV observatories, the VERITAS collaboration has developed different monitoring algorithms that are described in the following section.

\section{Data Analysis}
Three different analysis chains have been developed inside the VERITAS collaboration to monitor \FL data and promptly identify GeV flaring sources: daily light curves of selected blazars, an all-sky analysis to identify new flaring sources, and an algorithm that looks for clusters of high-energy photons in time and arrival direction. The three analysis pipelines automatically download \FL data and process it using the Fermi Science Tools v9r18 \cite{fssc}.

\subsection{Blazar light curves}
%The  analysis makes use of the Fermi Science Tools v9r18. 
The main goal of this analysis is to produce daily and weekly light curves for a list of blazars of interest in order to identify high flux states in the GeV band.

The software
consists of a series of Python scripts which are executed with two separate Perl wrappers.
The analysis chain consists of two main branches, described below.

The first branch of the analysis chain performs the time-averaged spectral analysis for a
given source over an energy range between 200\,MeV to the highest photon energy associated with
the source in question. 
For sources with low galactic latitudes, the photon data files (after data selection) are significantly larger than those of high galactic latitude sources. %(& 50MB versus  15 MB,
The memory usage for the unbinned likelihood analysis goes linearly with the
number of photons. As such, it is not possible, for some low-galactic-latitude sources, to
perform an unbinned likelihood analysis. If a photon data file has  a size $>50$\,MB after
event selection, the time-averaged spectral analysis is performed using a binned likelihood
analysis rather than the unbinned analysis.
Data within $15^{\circ}$ of the source are selected for the time-averaged spectral analysis. The
source model is built from the Fermi 1FGL Catalog and includes sources out to an angular
separation of $20^{\circ}$. Spectral parameters for sources beyond $15^{\circ}$ are fixed to their 1FGL values.
All model parameters within $15^{\circ}$ of the source of interest are initially left free. Then the model is fit to the data
%DRMNGB minimizer is used
 to find the general location of the minimum in model parameter
space. All model parameters beyond $10^{\circ}$ are then fixed to their best fit values. The data
is then refit
% using the MINUIT minimizer 
to determine the best fit model parameters, and
associated errors, for the source of interest.
The second branch of the analysis chain calculates the daily and weekly lightcurves for each
source. 
%Each lightcurve
%is calculated using data selected from 200\,MeV, and 1\,GeV, extending up to 300\,GeV. 
Lightcurves of the integral flux above 300\,MeV and 1\,GeV are produced.
The spectral index of the source of interest is fixed to the time-averaged value and only the source flux is fit. 
If a particular data
point in a lightcurve has a TS value less than 0.1, its flux and associated error are set to 0 to
improve the clarity of the lightcurves. Upper limits are not calculated here since they
are not used for triggering flares.
The fitting procedure is the same as that for the time-averaged analysis, with one additional step. While fitting over smaller time intervals, there will be sources in the model that
have no associated photons. This leads to an excessive number of degrees of freedom that can prevent
the fit from converging. To remedy this, after the fitting procedure is performed, all sources
with TS values less than 4 are removed from the model. The data is then refit. This process
is repeated until there are no longer sources with TS values less than the specified threshold.
It typically does not take more than two iterations to remove all non-significant sources.

\begin{figure*}[th]
  \centering
  \includegraphics[width=0.9\textwidth]{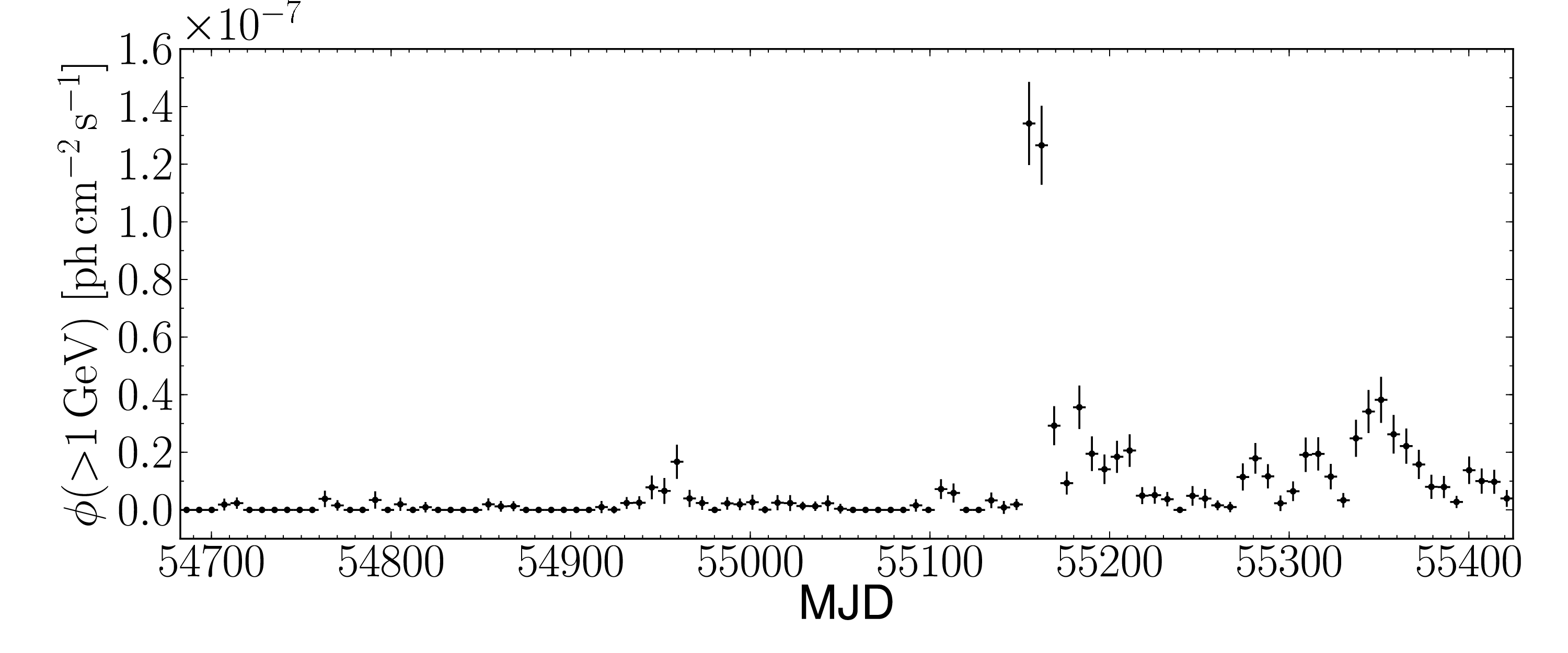}
  \includegraphics[width=0.9\textwidth]{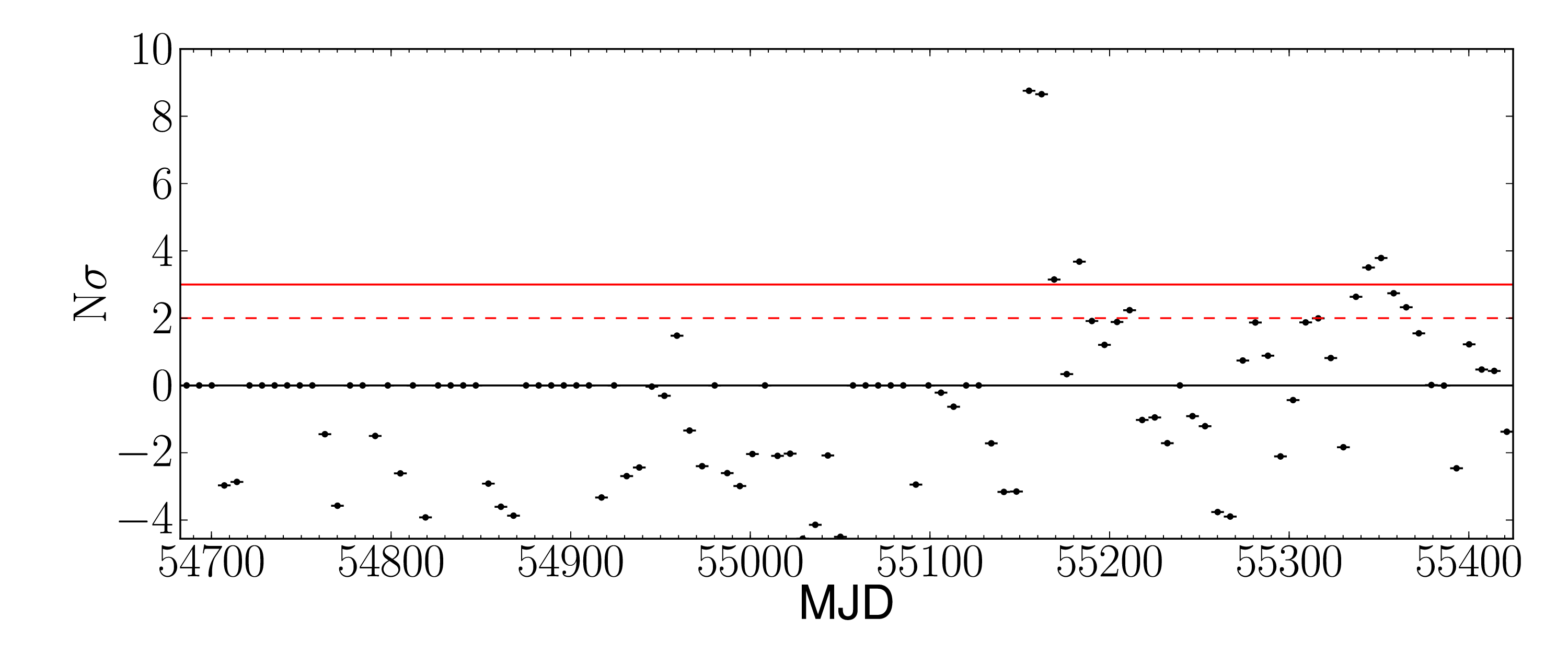}
  \caption{({\it top}) Gamma-ray light curve for the quasar CGRaBS~J1312+4828 in 1-week time bins. ({\it bottom}) Significance of the fluctuations of each flux point with respect to the source average flux. A dashed (solid) line indicates $2\sigma$ ($3\sigma$) fluctuations.
    }
  \label{fig:lc}
 \end{figure*}

To evaluate the significance of a particular flare, the significance ($S$) of a fluctuation in flux is evaluated against the average $\gamma$-ray flux from the source following the equation:
\begin{equation}
S = \frac{\phi_i - \phi_{avg}}{\sqrt{\sigma_i^2 + \sigma_{avg}^2}}
\end{equation}
as shown in Figure~\ref{fig:lc}.

\subsection{All-sky monitoring}
This analysis is based on monitoring the all-sky \FL data daily to watch for increased $\gamma$-ray activity above 1\,GeV from any sky direction. This is useful to detect $\gamma$-ray flares from unknown objects, particularly blazars and galactic transients.

\begin{figure*}[th]
\centering
\includegraphics[height=1.7in]{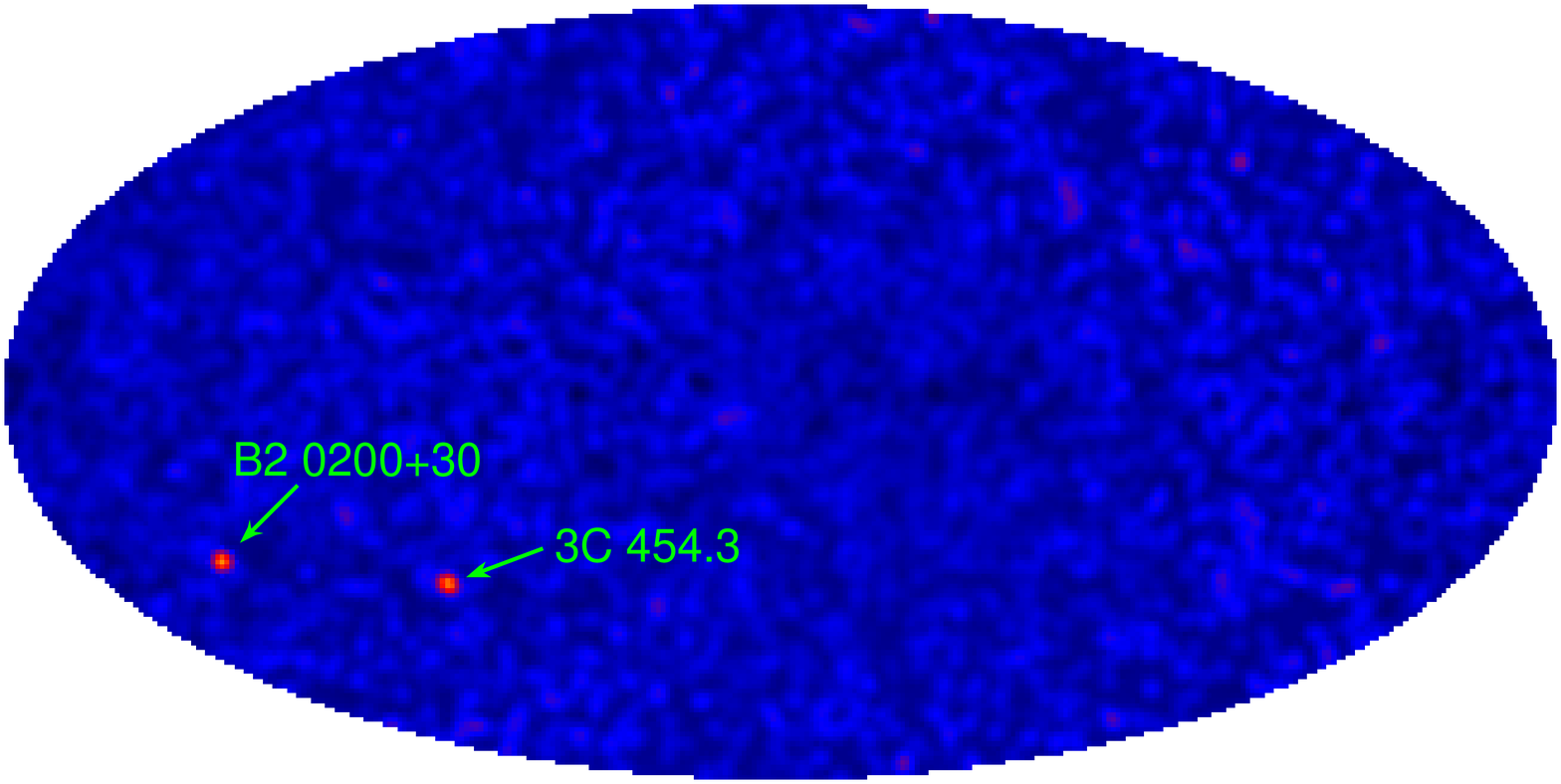}
\includegraphics[height=1.7in]{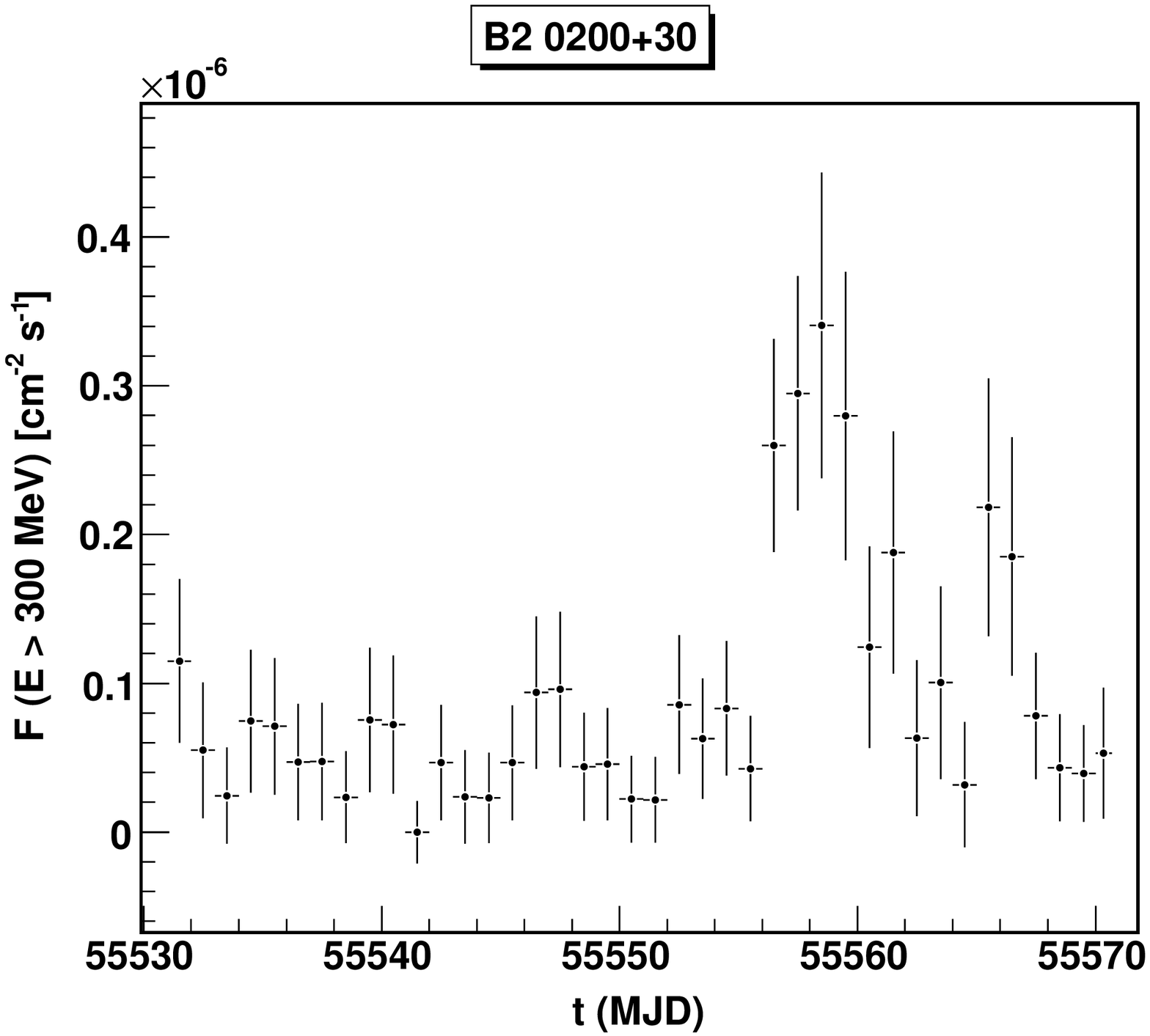}
\caption{ ({\it left}) \F variability map from 2010 December 29th showing the high flux states at $E>1$\,GeV from 3C~454.3 and B2~0200+30, which was observed by VERITAS after approval of a directors discretionary time request. ({\it right}) Daily light curve for $E>300$\,MeV measured by \F covering the $\gamma$-ray flare.}
\label{fig:b2}
\end{figure*}

For this analysis, an all-sky exposure corrected map ($1^{\circ} \times 1^{\circ}$ bins) is produced including all photons with $E > 1$\,GeV collected by \FL during the first 21 months of the mission. This map is taken as a static/averaged picture of the sky in \g rays. An equivalent all-sky map is created every day integrating data from the last 3 days. A variability parameter ($V$) is then calculated on a bin by bin basis:. 
\begin{equation}
V = \frac{\phi_{3 day}-\phi_{21month}}{\phi_{21month}}
\end{equation}
An example of the resulting maps can be seen in Fig~\ref{fig:b2}. High values of $V$ are seen as hotspots in the map, and indicate an increased $\gamma$-ray activity in a particular sky direction compared to the 21-month average flux recorded by {\em Fermi}. In addition to the 3 day integration map, similar maps are produced integrating over 1, 7 and 30 days. The different time scales are chosen to test a wide range of variability profiles, going from very fast flares to prolonged high states. Whenever a hotspot is identified on a variability map, a detailed likelihood analysis of the region is carried out, producing the best-fit parameters to the spectrum of the flaring source and a light curve to monitor its activity above 1\,GeV in the last 10 days %see Figure~\ref{fig:lc})
. This result from the spectral analysis helps to extrapolate the Fermi derived flux to TeV energies, and understand if the source is flaring at levels detectable by ground-based telescopes.

\subsection{Individual photon counting}
This analysis looks for clusters of photons with $E>3$\,GeV in the \FL data coming from the direction of a known $\gamma$-ray source within a short period of time. A cluster of high energy photons identifies a source that is in a high flux state and naturally selects hard spectrum sources that are potentially interesting for VERITAS observations. This analysis is particularly suited for sources that show rapid variability such as quasars.
%relies in the low $\gamma$-ray background that \FL has at energies . 

\begin{figure}[th]
\centering
\includegraphics[width=0.9\columnwidth]{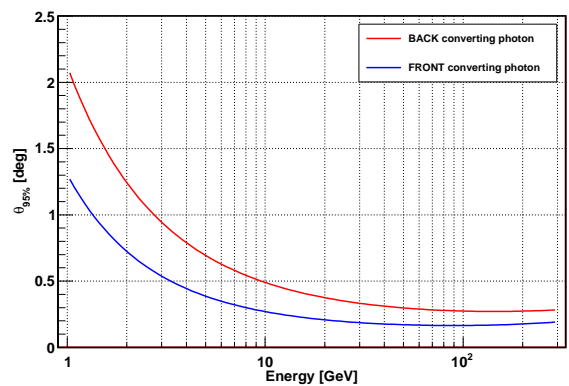}
\caption{Evaluated \F PSF for front and back converting photons as a function of energy.}
\label{fig:psf}
\end{figure}

\begin{figure*}[th]
\centering
\includegraphics[width=0.8\textwidth]{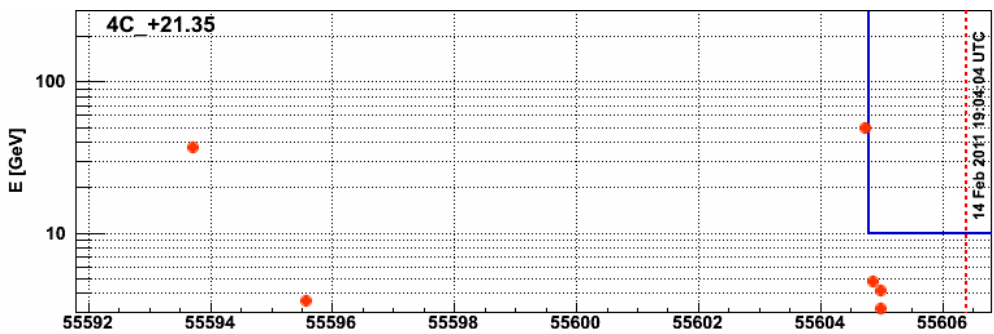}
\caption{Plot showing the arrival time and energy of photons coming from the direction of the quasar 4C~21.35. The blue box indicates a tentative trigger criterion. The cluster of high energy photons (orange dots) at MJD 55605 can be identified as high $\gamma$-ray activity coming from the source.}
\label{fig:photons}
\end{figure*}

The data processing is quite simple. Photons with $E>3$\,GeV are selected. Then the photon list is filtered, and only photons within the 95\% containment radius from the coordinates of the source of interest are kept. The containment radius is previously extracted from the P6\_V3\_DIFFUSE instrument response functions \cite{fssc} and is evaluated separately for front and back converting photons as a function of the photon energy (Fig~\ref{fig:psf}). The remaining photons, which are coming from the $\gamma$-ray source of interest with almost negligible background, are plotted as a function of arrival time and photon energy (Fig.~\ref{fig:photons}). The plots show the arrival time and energy (orange dots) of high energy photons for the selected sources last 15 days. Alerts are typically sent to VERITAS observers when several photons are recorded from a particular source within the last two days.

\section{Summary}
Three automatic analysis pipelines for \FL data have been presented. The different techniques allow the identification of $\gamma$-ray flares from known sources with high sensitivity, the detection of new flaring sources in any sky direction, and a prompt reaction to short-lived $\gamma$-ray flares from rapidly variable objects. The analysis presented here allows VERITAS to react in the shortest possible time to $\gamma$-ray flares detected by {\em Fermi}, maximizing the chances of a simultaneous detection at TeV energies.

\section{Acknowledgment}
This research is supported
by the NASA grant  NNX10AP66G. VERITAS is supported by grants from the US Department of Energy, the US National Science Foundation, and the
Smithsonian Institution, by NSERC in Canada, by Science Foundation Ireland, and by STFC in the UK. We acknowledge the
excellent work of the technical support staff at the FLWO and at the collaborating institutions in the construction and
operation of the instrument.

\clearpage

\end{document}